\documentclass[conference]{IEEEtran}
\IEEEoverridecommandlockouts
\usepackage{cite}
\usepackage{amsmath,amssymb,amsfonts}
\usepackage{graphicx}
\usepackage{textcomp}
\usepackage{xcolor}
\def\BibTeX{{\rm B\kern-.05em{\sc i\kern-.025em b}\kern-.08em
    T\kern-.1667em\lower.7ex\hbox{E}\kern-.125emX}}

\usepackage{bm}

\def\eqref#1{equation~\ref{#1}}

\def\1{\bm{1}}

\def\vg{{\bm{g}}}

\def\vx{{\bm{x}}}

\DeclareMathAlphabet{\mathsfit}{\encodingdefault}{\sfdefault}{m}{sl}
\SetMathAlphabet{\mathsfit}{bold}{\encodingdefault}{\sfdefault}{bx}{n}

\usepackage{array}
\usepackage{xspace}

\usepackage[normalem]{ulem}
\usepackage{cite}

\newcolumntype{L}[1]{>{\raggedright\let\newline\\\arraybackslash\hspace{0pt}}m{#1}}
\newcolumntype{C}[1]{>{\centering\let\newline\\\arraybackslash\hspace{0pt}}m{#1}}
\newcolumntype{R}[1]{>{\raggedleft\let\newline\\\arraybackslash\hspace{0pt}}m{#1}}

\makeatletter
\DeclareRobustCommand\onedot{\futurelet\@let@token\@onedot}
\def\@onedot{\ifx\@let@token.\else.\null\fi\xspace}

\makeatother

\newcommand{\congater}{\textsc{ConGater}\xspace}

\newcommand{\passt}{\textsc{PaSST}\xspace}

\usepackage{graphicx}
\graphicspath{ {./images/} }
\usepackage{amsmath,amssymb}
\usepackage{multirow}
\usepackage{caption}
\usepackage{subcaption}
\usepackage{algorithm} 
\usepackage{algpseudocode}
\usepackage{arydshln}
\usepackage{booktabs}
\usepackage{multirow}
\usepackage{fdsymbol}
\usepackage{siunitx}
\usepackage[export]{adjustbox}
\usepackage{url}

\begin{document}

\title{Domain Information Control at Inference Time \\ for Acoustic Scene Classification
\thanks{This work received financial support by the State of Upper Austria and the Federal Ministry of Education, Science, and Research through grant LIT-2021-YOU-215 and basic funding of the LIT AI Lab.}
}


\author{\IEEEauthorblockN{\textsuperscript{1,2}Shahed Masoudian, \textsuperscript{2}Khaled Koutini, \textsuperscript{1,2}Markus Schedl, \textsuperscript{1,2}Gerhard Widmer, \textsuperscript{1,2}Navid Rekabsaz}
\IEEEauthorblockA{\textit{\textsuperscript{1}Johannes Kepler University Linz, Institute of Computational Perception} \\
\textit{\textsuperscript{2}Linz Institute of Technology (LIT), AI Lab} \\
\textit{Austria}}}

\maketitle

\begin{abstract}

Domain shift is considered a challenge in machine learning as it causes significant degradation of model performance. In the Acoustic Scene Classification task (ASC), domain shift is mainly caused by different recording devices. Several studies have already targeted domain generalization to improve the performance of ASC models on unseen domains, such as new devices. Recently, the Controllable Gate Adapter (\congater) has been proposed in Natural Language Processing to address the biased training data problem. \congater allows controlling the debiasing process at inference time. \congater's main advantage is the \emph{continuous} and \emph{selective} debiasing of a trained model, during inference. In this work, we adapt \congater to the audio spectrogram transformer for an acoustic scene classification task. We show that \congater can be used to selectively adapt the learned representations to be invariant to device domain shifts such as recording devices. Our analysis shows that \congater can progressively remove device  information from the learned representations and improve the model generalization, especially under domain shift conditions (e.g. unseen devices). We show that information removal can be extended to both device and location domain. Finally, we demonstrate \congater's ability to enhance specific device performance without further training~\footnote{Source Code: \url{https://github.com/ShawMask/dcase22_congater}}.

\end{abstract}


\section{Introduction}


Domain Generalization is a critical topic in  Deep Neural Networks (DNN). The performance of conventional DNN methods drastically degrades under domain shift conditions when evaluated on new domains~\cite{wang2018deep}. 
Therefore, the ability of machine learning models to generalize to these new and unseen domains is crucial in real-world applications. Methods to improve the generalization of DNNs to a new domain are well studied in different fields, such as computer vision~\cite{venkateswara2020domain}, audio perception~\cite{kim2022domain} and natural language processing~\cite{jin2022deep}. 

This problem of domain generalization has drawn the attention of the Detection and Classification of Audio Events (DCASE) community, and datasets were constructed to test the generalization of common machine learning models under domain shift conditions~\cite{Martínmorato2022lowcomplexity}.

Convolutional Neural Networks (CNNs) have dominated the acoustic scene classification literature~\cite{kim2022domain, koutini2020cp}. More recently, vision transformers were adapted to audio tasks and shown to outperform CNNs on ASC tasks~\cite{koutini2021efficient}. Schmid et al.~\cite{Schmid2022} fine-tuned a transformer model pre-trained on a large dataset on an ASC dataset and used it as a teacher for knowledge distillation to low-complexity CNNs. Kim et al.~\cite{Lee2022} adapted a CNN model to this task.
In order to address the domain generalization problem, Schmid et al.~\cite{Schmid2022} used Frequency Mixstyle data augmentation to improve the performance on unseen devices. Kim et al.~\cite{kim2022domain} introduced ResNorm in BC-ResNet and Relaxed Instance Frequency Normalization (RFIN) as a new normalization method to achieve state-of-the-art results on unseen devices. Another approach to tackle the Domain Generalization issue is unsupervised Domain Adaptation (DA). Using Wasserstein distance, Drossos et al.~\cite{drossos2019unsupervised} learned domain invariant feature representations and improved the result of the ASC model on an unseen domain. Gharib et al.~\cite{gharib2018unsupervised} also used adversarial training to learn domain invariant representations to improve the performance of the ASC model on unseen devices. 

Invariant representation learning is also widely studied in the field of Natural Language Processing (NLP) on scenarios such as domain adaptation~\cite{jin2022deep}, domain transfer~\cite{ruder-etal-2019-transfer}, and mitigation of societal biases. In particular, the bias mitigation methods aim at removing sensitive information (e.g. Gender) from the DNN embeddings to improve fairness. The introduced approaches share many conceptual and methodological commonalities with domain invariant representation in DA, such as the utilization of adversarial techniques~\cite{ganin2016domain}. These approaches are widely studied in NLP literature~\cite{kashyap2022towards,kumar2023parameter,yuan2021improving}. Recent studies in this field focus on modular neural networks, where end-users can choose between debiased and biased models. These methods approach this by adding a separate module such as Adapters~\cite{houlsby_2019_param_efficient,lauscher_2021_sustainable}, or sparse subnetworks~\cite{hauzenberger2022parameter,meissner_2022_debias} to network. In this case, instead of training the whole model, these separate modules are trained to improve training efficiency and add a modular capability to the network~\cite{lauscher_2021_sustainable}.  More recently, Controllable Gate Adapter~\cite{masoudian2023controllable}  (\congater) expand the mentioned work by providing the ability for continuous sensitive information removal from the trained model.

\begin{figure}
\centering
\includegraphics[scale=0.6]{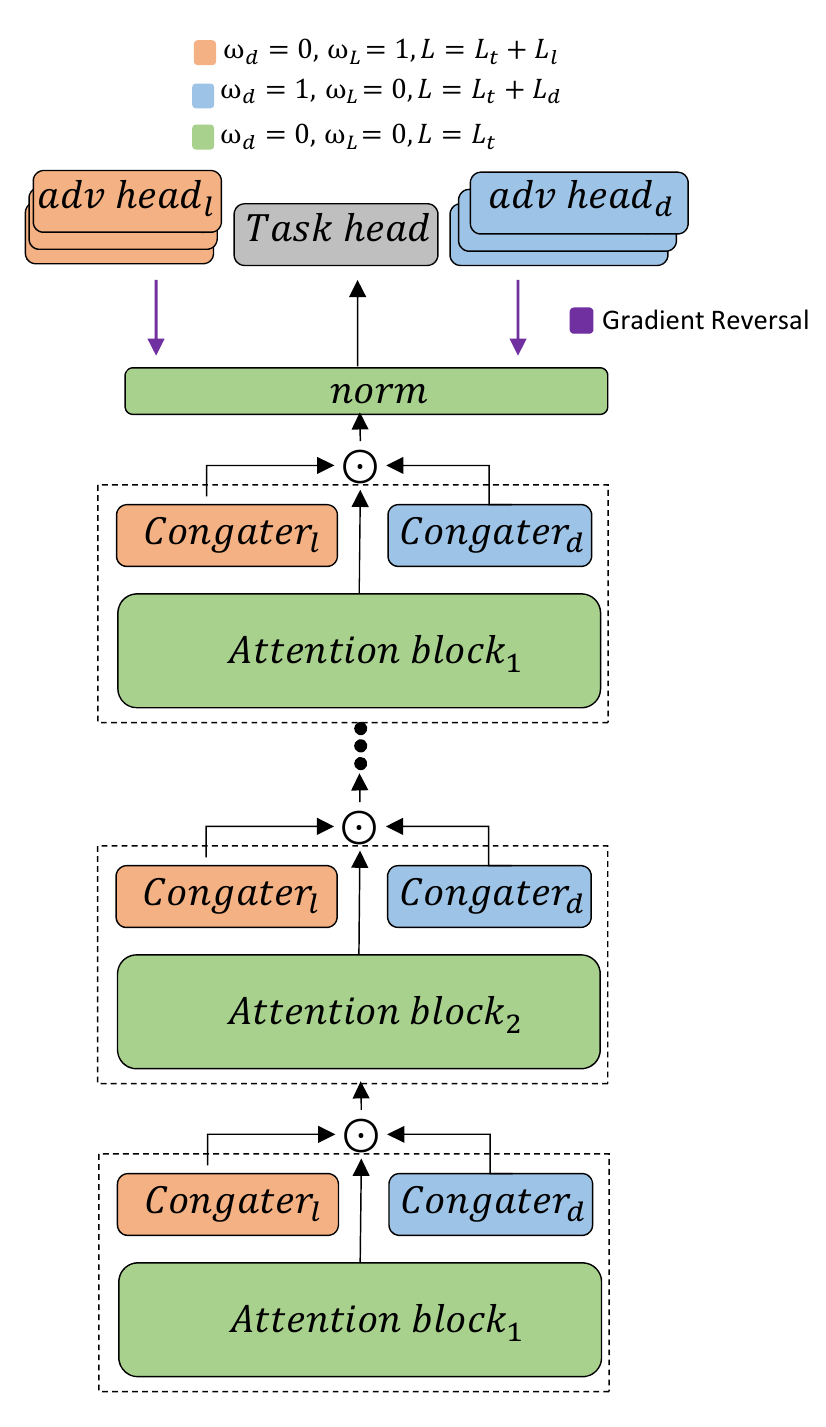}
\caption{Our proposed model using \congater modules, the location of the \congater layers, and the trainable parameters in each training step. The green blocks are trained with the loss of task ($L_t$), the blue blocks are trained adversarially to remove device information using task loss and reversed gradient loss of the device ($L_d$), the orange blocks are also trained adversarially to remove location information using task and reversed gradient location loss ($L_l$). For each domain, the average loss of the three adversarial heads is considered as domain loss. }
\label{fig:training_method}
\vspace{-7mm}
\end{figure}

In this work, we adopt the \congater idea from NLP and apply it to audio tasks. By adding \congater in between audio spectrogram transformer layers. We aim to remove device and location information to improve the generalization of audio spectrogram transformers on ASC tasks under domain shift conditions. We show that (1) \congater can effectively control the amount of information of device and location in the network on a continuous range; (2) by removing information from the device, we achieve on average 0.7\% higher overall accuracy compared to the baseline model, and 1.1\% accuracy improvement on unseen devices indicating the better generalization of the network; (3) our method can simultaneously be applied to both device and location domains; additionally, we demonstrate \congater's ability to fine-tune on a specific device by selecting suitable hyper-parameter at inference time without any need for further training.  

\section{Method}
\label{section:method}

For our experiments, we choose a transformer-based model \passt~\cite{koutini2021efficient} that performs very well on a wide range of audio tasks, including ASC tasks~\cite{schmid2022efficient}. \passt have a similar structure to Vision Transformers~\cite{dosovitskiy2020image}. \passt works by extracting patches of an input spectrogram and linearly projecting them to a higher dimension corresponding to the embedding size. The positional encoding---consisting of time and frequency positional encodings--- are added to these embeddings. Patchout~\cite{koutini2021efficient} is then applied to speed up the training and for regularization. The sequence is then passed through several self-attention layers and a feed-forward classifier. In our experiments, we use the pre-trained model on Audioset~\cite{audioset2017Gemmeke} and then fine-tune the model on the ASC task.
Since the \congater~\cite{masoudian2023controllable} was introduced as modules added to the transformer encoder layers of BERT~\cite{devlin2018bert} language model, we similarly extend \passt with \congater by adding the \congater modules after the attention blocks of \passt, as explained in the following.


\congater~\cite{masoudian2023controllable} extends the core idea of Adapter networks~\cite{houlsby_2019_param_efficient} by introducing a novel controllable gating mechanism, applied to input embeddings to deliver the desired learning objective such as learning a task or mitigating bias. Each \congater layer consists of a feed-forward network followed by a novel activation function called \emph{trajectory-sigmoid} (t-sigmoid). t-sigmoid is similar to normal sigmoid function, but with the extra sensitivity parameter
$\omega$, as formulated below:
 \begin{equation}
    \text{t-sigmoid}(v) = 1-\frac{\log_2{(\omega + 1)}}{1+e^{v}}\quad \omega \in [0,1]
\label{eq:t-sigmoid}
\end{equation}

The parameter $\omega$ can be manipulated manually at inference time and smoothly change the shape of the activation function. At $\omega=0$, the shape of t-sigmoid is the constant $y=1$ function. As $\omega$ increases, the activation function smoothly reshapes to a sigmoid function, adding a stronger non-linearity to the input.

In order to remove information of several domains, a \congater is dedicated to each domain and added to the network. As an instance, for the domain \textit{location}, the gating vector $\vg_{location}$ is achieved after applying the feed forward layer followed by t-sigmoid. The overall gate output is the element-wise multiplication of the vectors, which in our case is defined as $\vg=\vg_{device} \odot \vg_{location}$. The final output the layer is defined as the self-gate between the original input to \congater (namely the output of the attention layer denoted by $\vx$), and the final gating vector: $output = \vx \odot \vg$. Note that self-gate between any input and \congater of a domain with $\omega_{d}=0$ do not affect the input. By increasing $\omega_{d}$, the effect of removing this specific domain -- independent of other domain(s) -- increases. 
In our transformer-based architecture as shown in Figure~\ref{fig:training_method}, we add one \congater layer after each of the attention blocks.

\begin{figure*}[t]
\center
\vspace{-3mm}
\begin{subfigure}{0.32\textwidth}
\includegraphics[width=0.8\columnwidth,right]{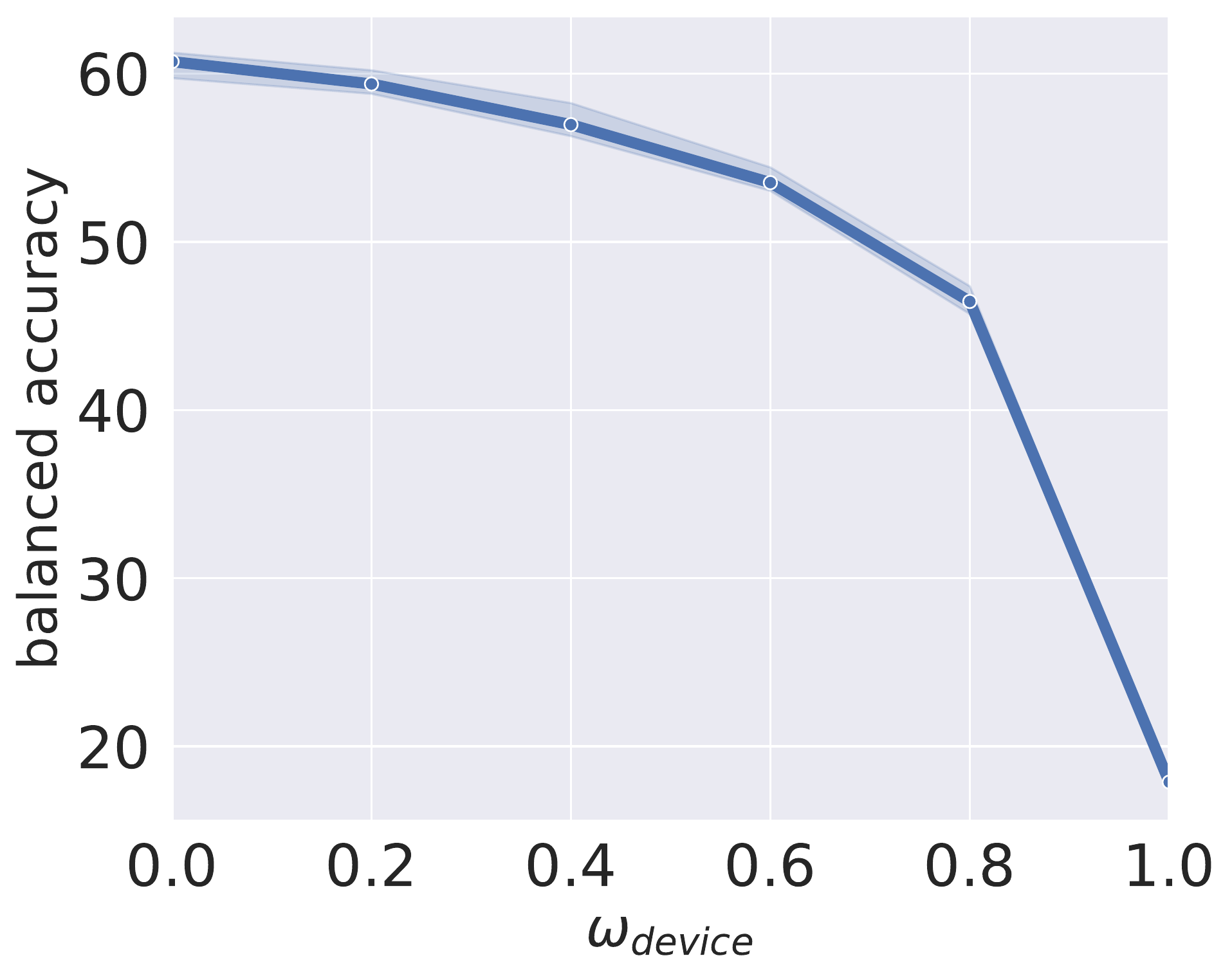}
\caption{Device probe balanced accuracy}
\label{fig:atk_device_acc_1d}
\end{subfigure}
\hfill
\begin{subfigure}{0.32\textwidth}
\includegraphics[width=0.85\columnwidth,right]{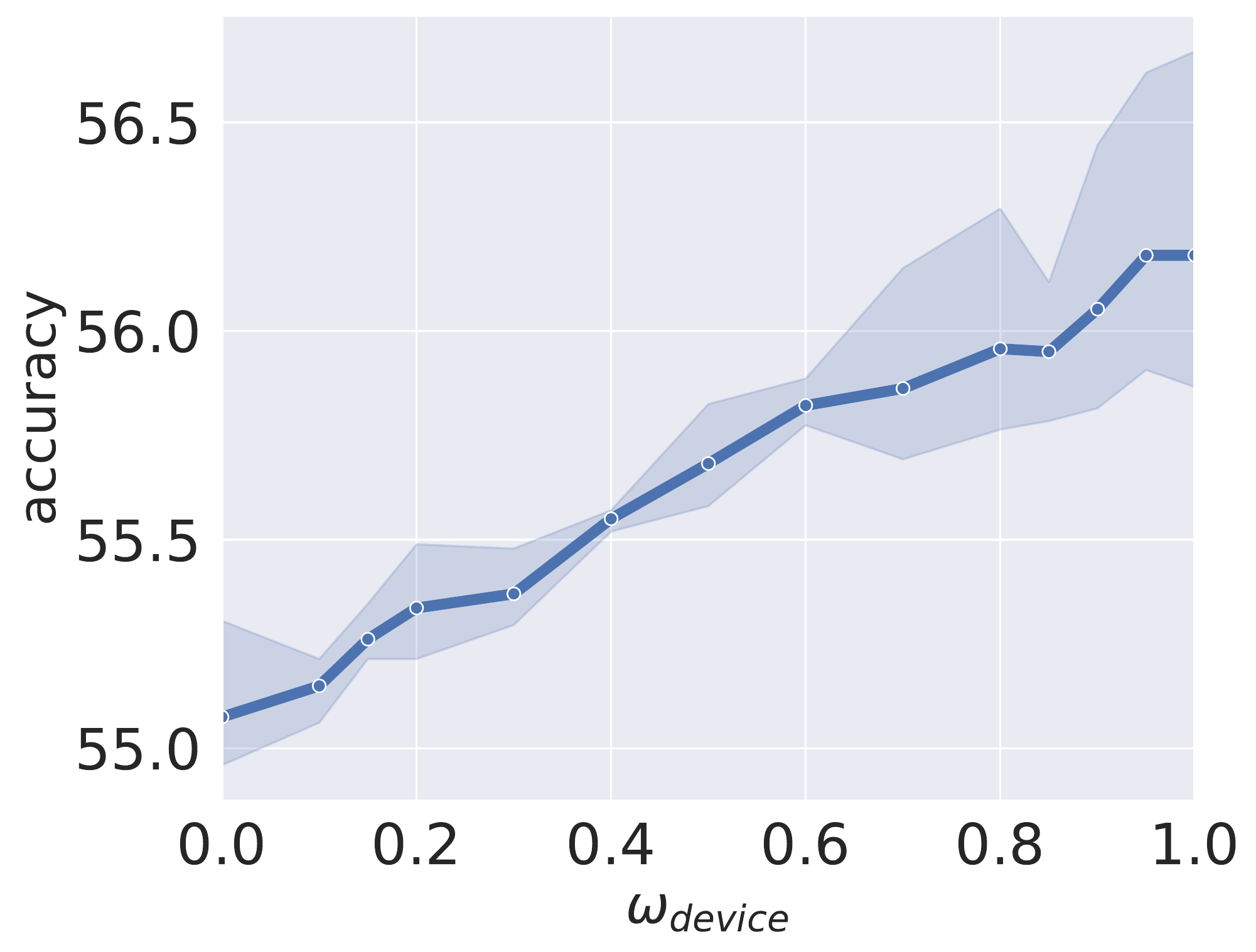}
\caption{Task accuracy on unseen devices}
\label{fig:unseen_acc_1d}
\end{subfigure}
\hfill
\begin{subfigure}{0.32\textwidth}
\includegraphics[width=0.85\columnwidth,right]{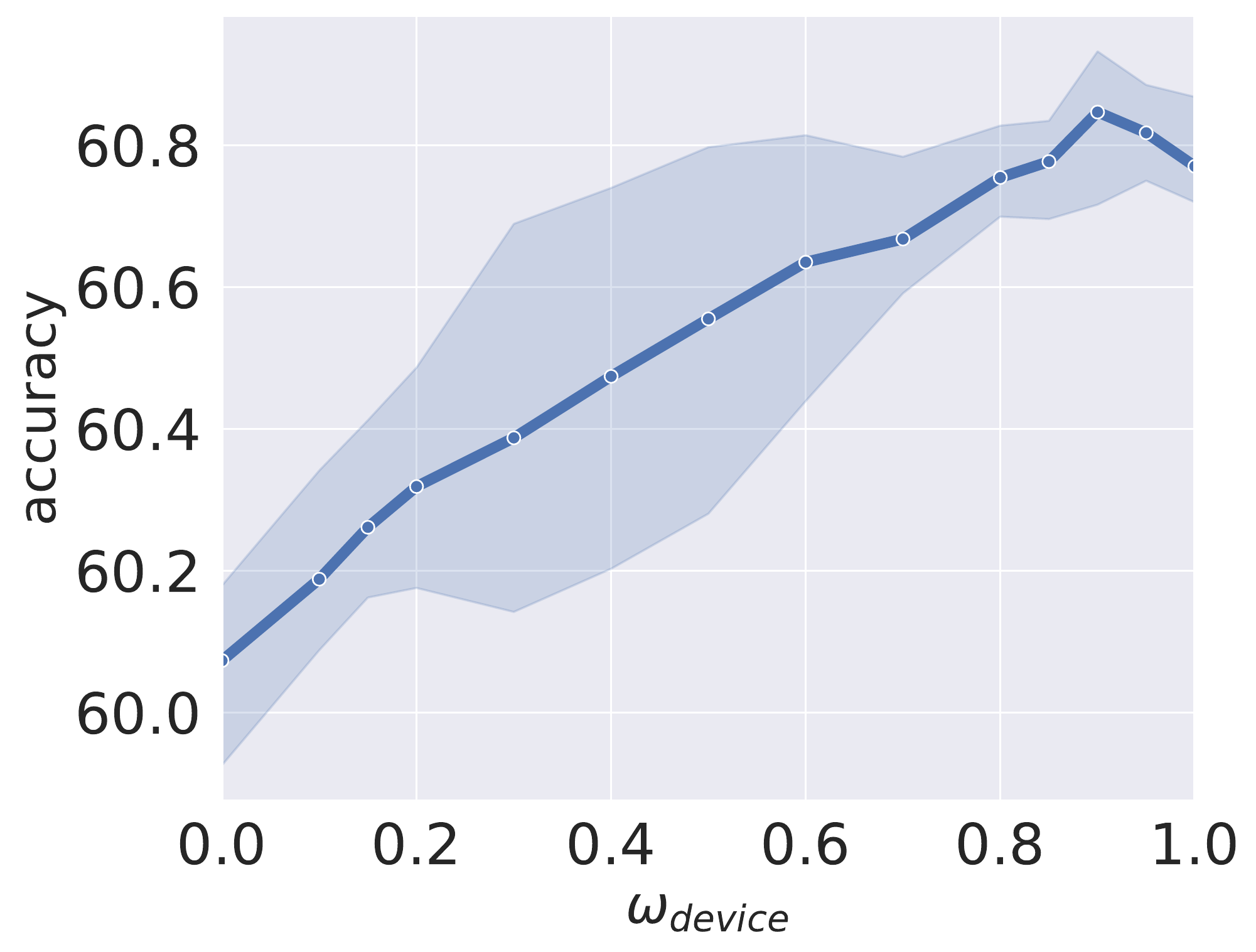}
\caption{Task accuracy on all devices}
\label{fig:overall_acc_1d}
\end{subfigure}

\caption{(a) Balanced accuracy of the device probe as we increase $\omega_{device}$ (b) model accuracy on unseen devices (S4-S6) as we increase only the $\omega_{device}$. (c) Overall accuracy of the model on validation set as we increase only $\omega_{device}$ 
}
\vspace{-6mm}
\end{figure*}


The training sequence for the proposed model consists of three steps, executed sequentially as each batch of training data arrives. The first step is \textbf{Task Training}, where parameters of the \passt in addition to the task-head are trained with $\omega_{device}=0$ and $\omega_{location}=0$. 
The second step is \textbf{Device Removal}, in which the parameters of the device \congater in addition to the task-head are trained with $\omega_{device}=1$ and $\omega_{location}=0$ to learn the task and remove the device-related information. Finally, in the third step \textbf{Location Removal}, the parameters of the location \congater in addition to the ones of the task-head are trained with $\omega_{device}=0$ and $\omega_{location}=1$ in order to learn the task and remove location-related information. We utilized Domain Adversarial Neural Network (DANN)~\cite{ganin2016domain} to remove information from embeddings. In this method, an additional sub-network (\textit{adversarial head}) is added to predict device/location labels from the original network embeddings. The reversed gradient loss is added to the network as an extra objective ($L_{total} = L_{task} + L_{adv}$). This objective--because of the gradient reversal--aims at removing the domain information  by maximizing the domain confusion in the original network embedding. 

\section{Experiment Setup}
\textbf{\textit{Dataset.}} The dataset used for our experiments is the adapted version of TAU Urban Acoustic Scenes 2022 Mobile development dataset used for the Acoustic Scene Classification (ASC) task~\cite{Martínmorato2022lowcomplexity} in the DCASE 2022 challenge.
The dataset contains 1-second audio recordings of 10 different acoustic scenes. The development set consists of audio recordings from 9 different devices, 3 real devices (A, B, C), and 6 simulated devices (S1-S6). The audios are recorded in 12 different European cities. We split 70\% of the development set for training in which only 6 devices (A, B, C, S1-S3) are used for training and the other 3 (S4-S6) are only seen during validation, which we refer to as \emph{unseen devices}. Similarly for the cities, during training 10 cities are available and 2 cities are available only during validation. 

\textbf{\textit{Preprocessing.}}  
We use a sampling rate of 32kHz. We apply Short Time Fourier Transformation (STFT) with a window size of 800 with an overlap of 320 (40$\%$) to generate the spectrograms. We apply mel-filters bank 128 in a similar setup to ~\cite{Schmid2022}. The final spectrograms have 128 mel-frequency bins and 100 time frames.

\begin{table*}[t]
\begin{tabular}{l|p{9mm}p{9mm}p{10mm}|p{9mm}p{9mm}p{9mm}p{9mm}p{9mm}p{10mm}||p{9mm}p{9mm}}
   \multirow{2}{*}{\textbf{Model}}  &  \multicolumn{3}{c|}{\textbf{Unseen Devices}}  & \multicolumn{6}{c||}{\textbf{Seen Devices}} & \multirow{2}{*}{\textbf{Unseen}}& \multirow{2}{*}{\textbf{Overall}} \\ 
   
   
    &  S4 & S5 & S6& A & B & C &  S1 & S2  & S3 & &  \\
     
      \midrule
Baseline &$56.9_{0.6}$&$57.3_{0.4}$&$50.9_{1.3}$
&$72.6_{0.3}$&$63.5_{0.5}$&$67.5_{0.4}$&$58.0_{0.9}$&$55.6_{0.8}$&$58.2_{0.5}$&$55.1_{0.2}$&$60.1_{0.1}$
\\                           
$\omega_d=0.6$, $\omega_l=0.0$ &$57.4_{0.5}$&$\mathbf{58.1_{0.4}}$&$51.9_{0.8}$&$72.7_{1.2}$&$63.5_{0.4}$&$67.7_{0.3}$& $59.4_{0.7}$&$56.2_{0.6}$&$58.7_{0.3}$&$55.8_{0.1}$&$60.6_{0.2}$\\
$\omega_d=0.7$, $\omega_l=0.0$ &$57.3_{0.4}$&$58.0_{0.2}$&$52.2_{1.0}$&$72.6_{1.1}$&$63.5_{0.6}$&$\mathbf{67.8_{0.3}}$&$59.5_{0.5}$&$56.2_{0.6}$&$58.8_{0.3}$&$55.9_{0.3}$&$60.7_{0.1}$\\
$\omega_d=0.9$, $\omega_l=0.0$ &$57.5_{0.3}$&$57.9_{0.4}$&$52.7_{0.9}$&$72.8_{0.7}$&$63.6_{0.2}$&$67.5_{0.4}$&$\mathbf{59.7_{0.6}}$&$\mathbf{56.6_{0.7}}$&$59.2_{0.6}$&$56.1_{0.3}$&$\mathbf{60.8_{0.1}}$\\
$\omega_d=0.9$, $\omega_l=0.1$ &$57.4_{0.2}$&$58.1_{0.5}$&$52.7_{1.0}$&$72.7_{0.7}$&$\mathbf{63.8_{0.6}}$&$67.6_{0.4}$& $59.5_{1.0}$&$56.3_{0.4}$&$59.0_{0.4}$&$56.1_{0.5}$ &$60.8_{0.3}$ \\
$\omega_d=1.0$, $\omega_l=0.1$ &$\mathbf{57.7_{0.3}}$ & $58.1_{0.7}$ & $\mathbf{53.0_{1.0}}$&$\mathbf{72.8_{0.3}}$ & $63.7_{0.7}$ & $67.3_{0.5}$&$59.4_{0.9}$ & $56.1_{0.4}$ & $\mathbf{59.3_{0.6}}$&$\mathbf{56.3_{0.6}}$&$60.8_{0.3}$\\
\midrule
Device-specific Tuning &$\mathbf{57.9_{0.3}}$ & $\mathbf{58.1_{0.4}}$ & $\mathbf{53.0_{1.0}}$&$\mathbf{72.8_{0.3}}$ & $\mathbf{63.8_{0.6}}$ & $\mathbf{67.8_{0.3}}$&$\mathbf{59.7_{0.6}}$ & $\mathbf{56.6_{0.7}}$ & $\mathbf{59.3_{0.6}}$&$\mathbf{56.3_{0.6}}$&-\\
achieved in [$\omega_d$,$\omega_l$] & [1.0,0.1] & [0.6,0.] & [1.0,0.1]& [1.0,0.1] & [0.9,0.1] & [0.7,0.] & [0.9,0.0] & [0.9,0.0] & [1.0,0.1] & [1.0,0.1] &-
\end{tabular}
\caption{
Comparison of the performance of our \congater-based model on target devices under various values of the $\omega$ parameters. We group devices as \textbf{Seen}, and \textbf{Unseen}, indicating whether they exist during training or only appear in evaluation. The subscripts $d$ and $l$ refer to device and location, respectively. Baseline is the case with $\omega_d=\omega_l=0.0$, and Device-specific Tuning refers to dynamically find the best performing $\omega$ combination for a specific target device.
}
\label{tab:congater}
\vspace{-0.5cm}
\end{table*}

\textbf{\textit{Training.}} We train each model with a batch size of 100 with \textit{CrossEntropy} loss, \textit{Adamw} optimizer, weight decay of 0.001, max learning rate of \num{1e-5} for \passt model, and \num{1e-4} for the \congater layers. We use a pre-trained \passt model on Audioset~\cite{audioset2017Gemmeke}, We initialize all \congater layers with zero weight and bias $b=5$. This initialization forces the output of \congater layers to 1 at the beginning of training. Each model is trained for 25 epochs with the following learning rate scheduler: for the first 3 epochs, the learning rate is exponentially increasing to its max value, in the next 3 epochs the model is trained with the max learning rate, and finally for the next 10 consecutive epochs the learning rate decreases to reach 0.01 of the max value. Adversarial training is used for \congater layers to remove device/location information. In order to have a smooth adversarial gradient flow during training, we use average the loss of three adversarial heads.
No normalization, augmentation, or dataset balancing is applied, in order to be able to check the direct effect of information removal on the baseline model where $\omega_{device} = 0$ and $\omega_{location} = 0$.

In this section, we discuss and analyze the results of our experiments. As mentioned in the previous sections, we train our proposed model, and the two independent \congater modules are trained to remove the domain information of the device and location. 
More concretely, at inference time each \congater parameter $\omega$ is used to remove the corresponding domain information from the embeddings of the network by increasing its value from zero to one.
The parameters $\omega_{device}$, $\omega_{location}$ control the intensity of the domain adaptation process for the device and the location, respectively. In our experiments, we mainly focus on the effect of $\omega_{device}$ on the network, namely in terms of device information leakage, and model performance evaluated on all and unseen devices. In addition, we report and analyze the results when controlling simultaneously over both device and location domains. To account for the possible variations in the results, we repeat each experiment three times and report the average and standard deviation of the evaluation results. 

\subsection{Device Information}

In order to check whether domain information is indeed removed from the network, we evaluate the degree of information leakage using additional classifier heads. This method is commonly referred to as \emph{probing} in the NLP literature~\cite{ kumar2023parameter}. Each probe is a two-layer feed-forward network with a ReLU activation function and is trained on the output embedding of the last attention layer of the network, before the task head. The probe aims to classify device labels from the embeddings. We retrain the probe for each value of $\omega_{device}$. The probes are trained for 5 epochs with a learning rate of \num{1e-4}. 
We calculate the balanced accuracy of the device classification on the validation set and report the average and standard deviation of the results attained from three independent runs.\\
Figure~\ref{fig:atk_device_acc_1d} shows that at $\omega_{device}=0$, namely the Baseline with no information removal from the model, the average balanced accuracy of the probe for device label is 60.7\%. This indicates that the embeddings of the model are highly informative about the recording device of the incoming audio. As we increase the $\omega_{device}$ and retrain the probes, the balanced accuracy consistently drops, indicating that the device information is decreasing in the embeddings. At $\omega_{device}=1$, we reach the lowest balanced accuracy of 17.9\%, showing that the \congater module has successfully removed most of the information about the device, as -- in contrast to the Baseline -- the probe is not able to predict device labels with high accuracy. Our results show that \congater can effectively remove domain information from the model, and the amount of information removal can be decided at inference time over a continuous range.

\subsection{Model Performance}

The main purpose of domain adaptation in ASC is to achieve better performance on unseen domains by generating domain invariant feature representation. We now turn our attention to the task accuracy of the model on unseen devices (S4-S6). Figure~\ref{fig:unseen_acc_1d} shows the task accuracy results of the network on unseen devices as we increase $\omega_{device}$. We observe that from the Baseline position $\omega_{device}=0$ (where no information from the device is removed), the average task accuracy on unseen devices starts at 55.1\%. By increasing $\omega_{device}$ and removing the device information, the accuracy of the network on unseen devices continuously improves and reaches its max at 56.2\% when $\omega=1$. The results on unseen accuracy indicate that increasing $\omega$ and removing more device information leads to improving model generalization across unseen devices. We also evaluate the overall (all devices) performance of the model with the same procedure as before. Looking at Figure~\ref{fig:overall_acc_1d}, we see that by increasing $\omega_{device}$, task accuracy increases and peaks at $\omega_{device}=0.9$ where the average task accuracy is 60.8\%, followed by a negligible drop at $\omega_{device}=1$. All in all the results are a clear indication that by removing device information from the network we achieve less informative embeddings which leads to better generalization on unseen devices as well as overall model performance.\footnote{Extensive 2D analysis of the domains is available in the GitHub repository}


\begin{figure}
\centering
\begin{subfigure}{0.34\textwidth}
\includegraphics[width=.9\columnwidth,center]{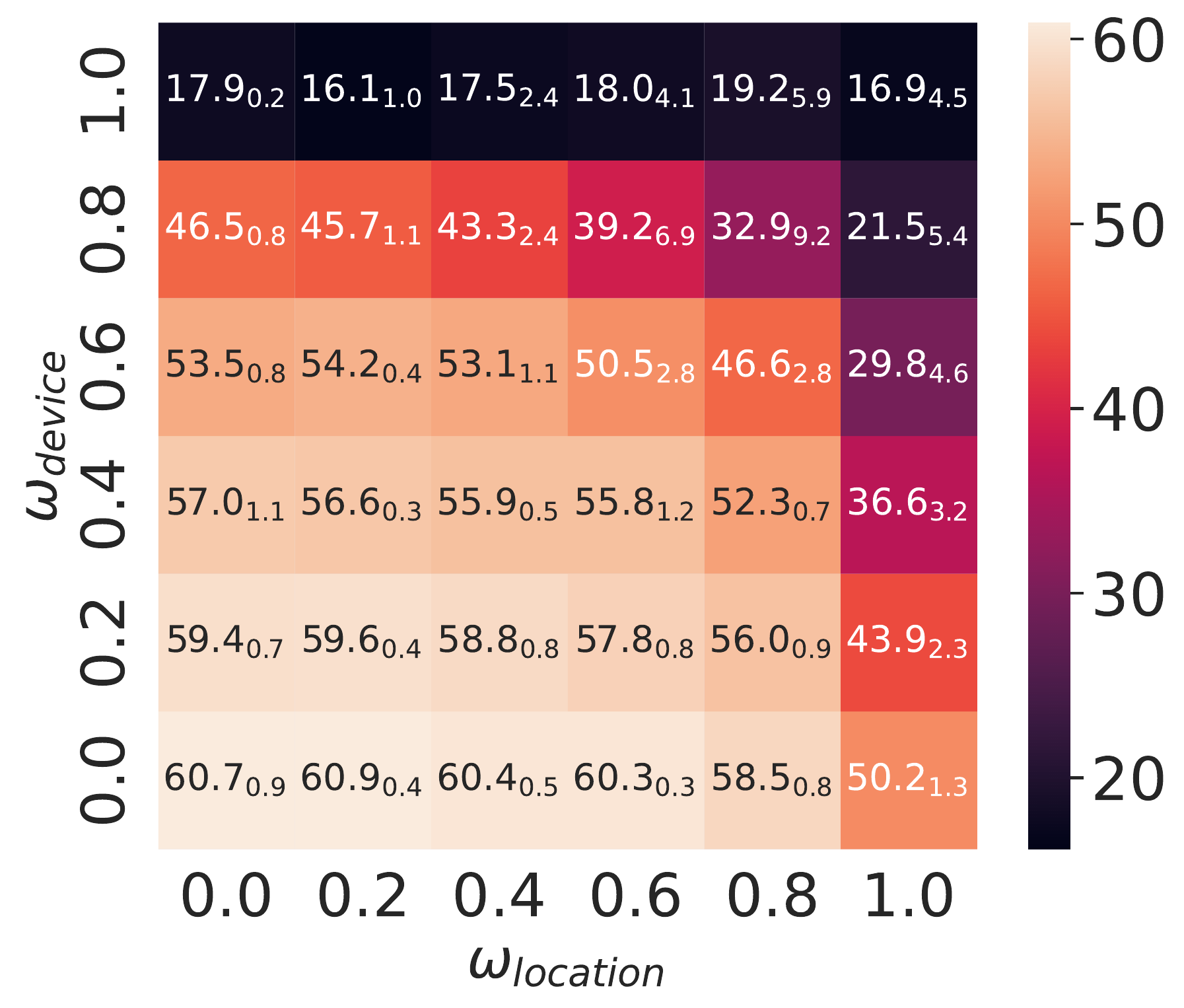}
\caption{Device probing}
\label{fig:atk_device_acc_2d}
\end{subfigure}
\begin{subfigure}{0.34\textwidth}
\includegraphics[width=.9\columnwidth,center]{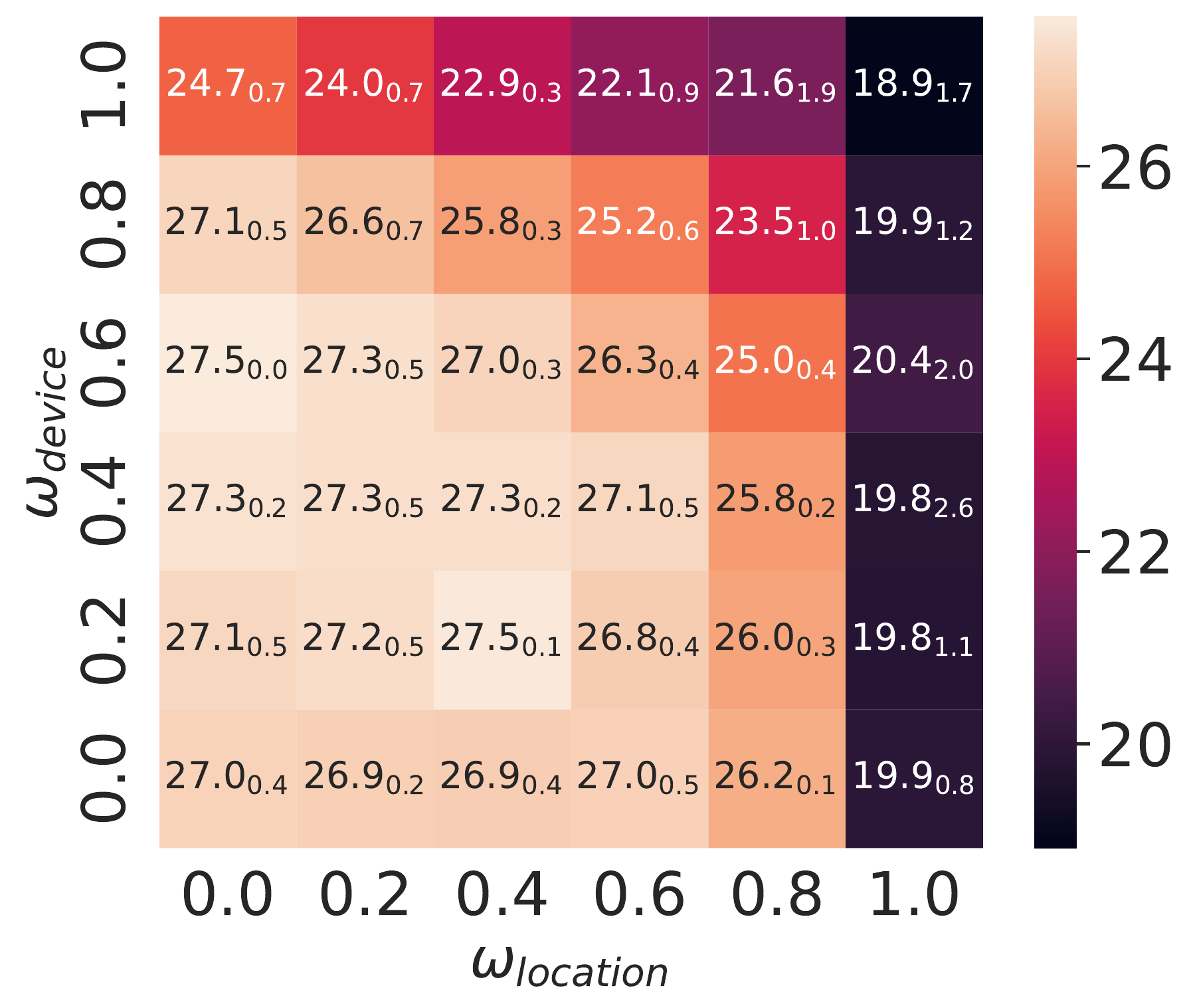}
\caption{Location probing}
\label{fig:atk_location_acc_2d}
\end{subfigure}
\caption{Balanced accuracy of the probes, measuring the information leakage of device (a) and location (b) in our model, when increasing device and location $\omega$ parameters.}
\label{fig:atk_location_bacc_2d}
\vspace{-7mm}
\end{figure}

\subsection{Device Specific Tuning}

In this section, we discuss the benefits of controllable information removal at inference, particularly in the scenario of finding the optimal model configuration according to a specific target domain. 
As shown in Figure~\ref{fig:atk_location_bacc_2d} and similar to previous experiments (Figure~\ref{fig:atk_device_acc_1d}), \congater-based model can adjust the information of both domains without additional training. Interestingly, the simultaneous removal of both factors on the top-right of the plot ($\omega_{device}=1$ and $\omega_{location}$=1) leads to a less information leakage of both device and location domains than information removal of one domain. 

Despite the successful removal of domain-specific information, our evaluation of task accuracy shows that removing only the location (in contrast to the device) does not result in meaningful performance improvement. However, \congater's ability to manipulate domain information can be leveraged to enhance model performance for a specific domain. We observed that by careful selection of the $\omega_{device}$ and $\omega_{location}$, we can improve model's performance with regard to one or several devices. This ability can be particularly advantageous when we introduce a new unknown device with a small subset of audio. Instead of re-training the whole model for a new device, we can simply select the suitable $\omega$ for the same pre-trained model at inference without any need for further training. The selection process can be either based on expert knowledge of the source and the target domain or by empirically validating model performance on a small subset of the target domain at inference (Grid search of the \congater parameter $\omega$). 
Table~\ref{tab:congater} shows the specific values of $\omega_{device}$, $\omega_{location}$ and model performance on each device. As it can be seen from the table there are unique $\omega_{device}$ and $\omega_{location}$ which result in better performance of the same model for one particular device. For instance, even though the best $\omega$ value for unseen devices is $\omega_{device}=1$, $\omega_{location}=0.1$, we can see that for device S2, model has 0.5\% better performance on average at $\omega_{device}=0.9$, $\omega_{location}=0.0$.  Note that with device-specific tuning, it is possible to optimize for a specific device, albeit at the expense of the performance of other devices.

\section{Conclusion}

In this paper, we adopt \congater from the NLP domain and implement it on audio spectrogram transformers. Training \congater on ASC dataset for multi-device domain adaptation, we observe that our model using \congater modules is effective at continuous device/location information removal from the network embedding at inference time. We observe that by increasing the sensitivity of the parameter $\omega_{device}$, the embeddings of the network effectively lose device information (domain labels), shown by the decrease in the balanced accuracy of the probes. We observe significant improvement in both unseen devices and overall accuracy for an already trained model by adjusting $\omega_{device}$ from 0 to 1. This observation indicates that removing device information leads to a more robust embedding for unseen devices. We observe that removing information from location alone does not significantly improve task accuracy, nor the unseen accuracy of the devices. Finally, we demonstrate that correct selection of \congater hyper-parameter for location and device leads to device-specific performance improvement.

\bibliographystyle{IEEEtran}
\bibliography{ref}


\end{document}